\documentclass[11pt,twoside]{article}
\usepackage{asp2006}
\usepackage{graphicx}

\begin{document}
\title{Infall and accretion} 
\author{F. Combes} 
\affil{LERMA, Observatoire de Paris, 61 Av. de l'Observatoire, F-75014, Paris, France}  

\begin{abstract}
  Gas infall and accretion play a fundamental role in galaxy
formation, and several processes of accretion are reviewed.
In particular the cold accretion may solve to some extent the
angular momentum problem in disk formation, while it is aggravated
by mergers. Gas accretion is one of the main actor in secular
evolution: it is required
to account for recurrent bar formation, and to explain the feedback
cycles of formation of bulges and black holes, with correlated masses.
Infall is also required to fuel a regular and almost stationary star
formation history. Star formation is quenched for galaxy in clusters when
gas accretion is suppressed through stripping. The central brighter
central galaxy can benefit however of gas accretion through cooling
flows, moderated by AGN feedback. Hot and cold feedback scenarios 
can be considered, to account for a stationary cooling flow, and explain
the filamentary CO and H$\alpha$ observed structures.
\end{abstract}

\section{Cold gas accretion on galaxies}

\begin{figure}[ht]
\begin{center}
\includegraphics[width=11cm]{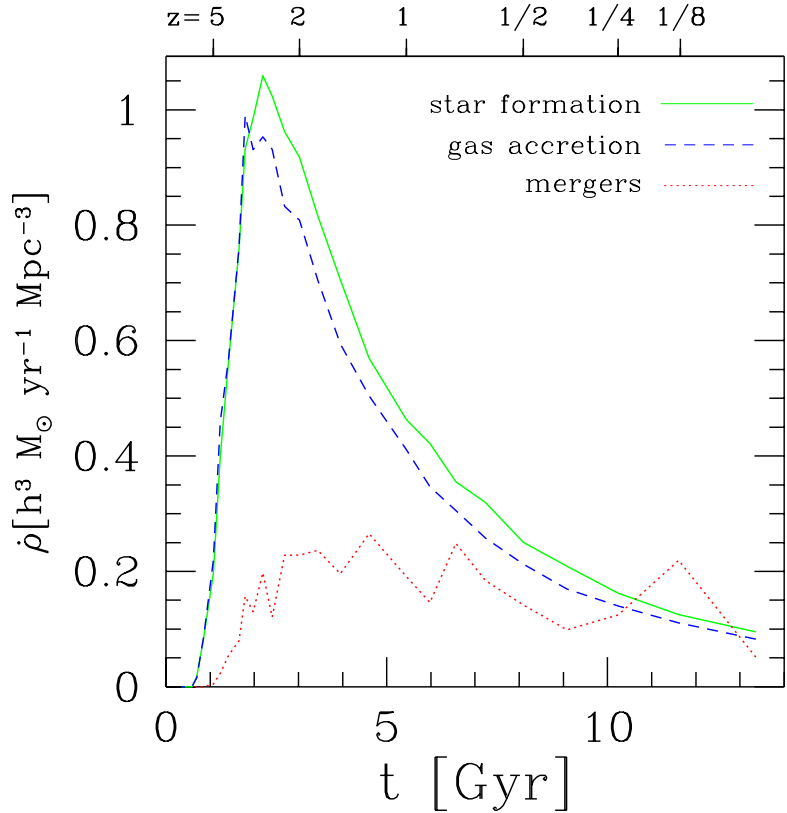}
\end{center}
\caption{Comparison as a function of time of the 
star formation rate (solid line, green), the smooth gas accretion rate onto
simulated galaxies (dashed line, blue), and mass gained in mergers
(dotted line, red), by Keres et al. (2005).}
\label{fig1}
\end{figure}

According to the conventional scenario of the $\Lambda$CDM
hierarchical galaxy formation, mass is assembled by mergers
essentially, and the baryonic gas is shock heated to the virial temperature
of the total structure ($10^6$ K for a Milky Way-type galaxy), before 
being accreted, as far as the hot gas cools down. This scenario 
ignores the possibility of dense gas present in contrasted structures, that
can cool much quicker, and radiate its gravitational energy while
remaining cold. Numerical simulations with enough resolution 
show now clearly two modes of accretion (Keres et al. 2005): 
hot gas accretion, mainly 
around massive structures, where shocks can be violent, and 
cold gas infalling along filaments, the fraction of cold gas being 
larger in low-mass haloes  (M$_{CDM} < 3 10^{11} M_\odot$).
The cold gas accretion dominates at high redshift. In addition,
the star formation rate history follows intimately the total
gas accretion, as shown in Figure \ref{fig1}.

The observation of a clear bimodality between galaxies -- a blue sequence still accreting gas and forming stars, 
while the red sequence consists of massive red galaxies, in general in rich environment, where gas and new stars are absent --
has been interpreted by the existence of these two modes of gas accretion, by 
Dekel \& Birnboim (2006). Below a certain halo mass of $\sim 10^{12} M_\odot$, the accretion shock is not 
sufficient to heat the gas to the virial temperature, and galaxy disks can be formed by cold gas flows, yielding 
efficient star formation. While in more massive structures, the gas heated by shocks is diluted, and sensitive to
 feedback due to active nuclei, quenching star formation, and producing "red and dead" galaxies, which today 
do not evolve any more, their observed properties changing only as expected for passive evolution.

The bimodality is observed on big samples of galaxies (SDSS, 2dF, MGC..), and separates clearly the early-type 
"red" galaxies from the late-type "blue" ones, according to their stellar distribution (Sersic index, for example), 
or their surface density at effective radius. The total galaxy mass is related to this separation, the red sequence 
being more massive (e.g. Driver et al. 2006).
 This observation is also reflecting the downsizing phenomenon: 
 early-type galaxies appear to have formed the majority of their stars very early in the universe,  while star 
formation is going on now in smaller 
spirals and dwarfs. Is this incompatible with the
hierarchical scenario?

\section{Star formation history and surface density}

 In fact, the two sequences of galaxies are distinguished by their
stellar surface densities (Kauffmann et al. 2003). Low Surface Density
(LSB) galaxies are small spirals or dwarfs, with
high gas content, high and young star formation, while High Surface Density
(HSB) are massive early-type galaxies, with concentrated mass, characterised
by an old stellar population. The transition occurs at the visible 
mass of $3 10^{10} M_\odot$, or stellar surface density of 
3 10$^8$ M$_\odot$/kpc$^2$. But star formation histories are more 
strongly correlated with surface mass density than with stellar mass.

This observation could be interpreted as the consequence of star formation feedback. At low mass 
concentration, the potential well is not deep enough to retain the gas after supernovae explosion, and star
 formation is quenched, which leads to a lower SF efficiency in LSB galaxies. Following Dekel \& Silk (1986), 
the energy provided by supernovae can be estimated as
E$_{SN} \sim \epsilon \nu$ dM*/dt  t$_{rad}$, where $\epsilon$ = 10$^{51}$ erg is the energy released by
 one supernova, the frequency of supernovae per unit mass is $\nu$ =1 for 100 M$_\odot$ stars formed, and
 the characteristic time t$_{rad}$ marks the end of the adiabatic phase. This amount of 
energy from supernovae is able to disperse the gas (of mass M$_g$), 
when it equals its kinetic energy 1/2 M$_g$ V$^2$, where V is the rotational velocity of the galaxy.
 There is a transition in galaxy rotational velocity (or potential well depth) 
where the gas begins to outflow, at the critical
velocity V$_{SN} \sim$ 100 km/s, which corresponds in order of magnitude 
to the transition found between LSB and HSB.

\subsection{Quenching of star formation}

\begin{figure}[ht]
\begin{center}
\includegraphics[width=10cm]{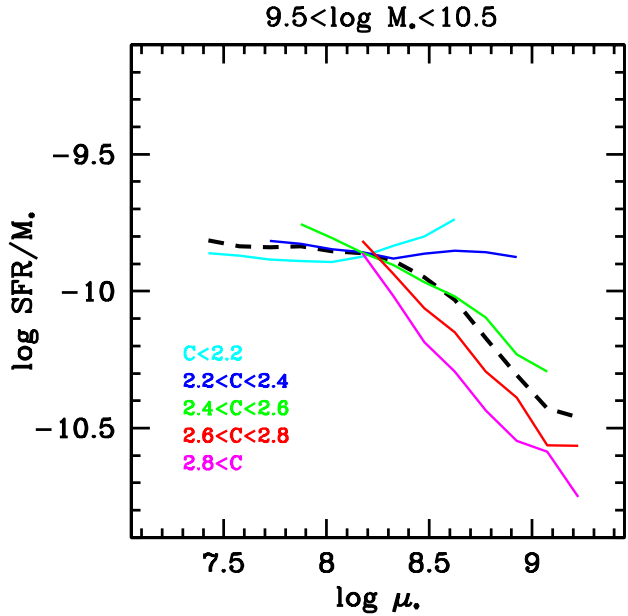}
\end{center}
\caption{The star formation rate per unit stellar mass rate depends essentially on surface density, and on 
the concentration. Here this specific SFR is plotted versus surface density for masses between 3 10$^9$ and 
3 10$^{10}$ M$_\odot$, and different concentrations in different colours (the black dashed line is the total),
 from Kauffmann et al. (2006).}
\label{fig2}
\end{figure}

As shown in Figure \ref{fig2}, the main parameter determining the star formation efficiency is the stellar 
surface density. Below the critical value of
3  10$^8$ M$_\odot$/kpc$^2$, the average star formation rate per unit mass does not depend on mass or 
surface density, nor concentration. The scatter in star formation rate increases with surface density.  
The SFR is then more bursty, and depends essentially on the external gas accretion. Above the threshold in 
surface density, the galaxy growth through star formation shuts down. The time-scale to consume the
 gas is inversely proportional to the stellar surface density.

\subsection{Origin of the bimodality}

There is therefore two interpretations for the origin of the bimodality: either the total mass is the determining
 factor, and the transition occurs at the critical halo mass of 3 10$^{11}$ M$_\odot$: above this halo mass, the 
external gas is not accreted cold, but is heated in shocks and then 
has no time to cool (subject in addition to AGN feedback, 
Dekel \& Birnboim 2006);
or above the critical surface density of stars of 3 10$^8$ M$_\odot$/kpc$^2$,
the gas is quickly transformed into stars, and the time spent in the
 "blue" star-forming sequence is short. In this view, galaxies pass from the red sequence to the blue sequence 
and back, but the blue phase is much shorter
at high mass and concentration than at low mass (Kauffmann et al. 2006).

\section{Gaseous filaments around Galaxies}

The numerical techniques are progressively refined to take into account
colder gas, and simulate the cold gas infall. While still most simulations consider the ``cold gas'' to be
at $10^4$K only, infalling multi-phase gas has been studied by 
T. Kaufmann et al. (2006), and the cooling/heating of dense clumps is now introduced in the SPH scheme, 
through considerations of the H$_2$ and HD
molecules (Maio et al. 2007).

The gaseous filaments are hard to observe, because of their low density and dilution. However, with the high 
sensitivity of present day deep searches, it is possible to see them at their connection to galaxies. The example 
of the edge-on galaxy NGC 891 is spectacular (Fraternali et al. 2007). Not only a thick gaseous envelope is seen
 at a thickness of 8kpc, but HI gas filaments are obvious up to 20kpc height above the galaxy plane. The rotational velocity of this gas 
decreases with z, and its origin cannot be due to
a galactic fountain phenomenon, coming from the star formation feedback, 
which will conserve angular momentum.

In some galaxies, the fountain effect is detected in addition
(for example the gas outflows in NGC 2403), while in general inflow is observed, like high velocity clouds 
(HVC, IVC) in the Milky Way.
Gaseous haloes are only explained by accretion of external gas
(Fraternali \& Binney 2006).

\section{Gas angular momentum lost in mergers}

\begin{figure}[ht]
\begin{center}
\includegraphics[width=10cm]{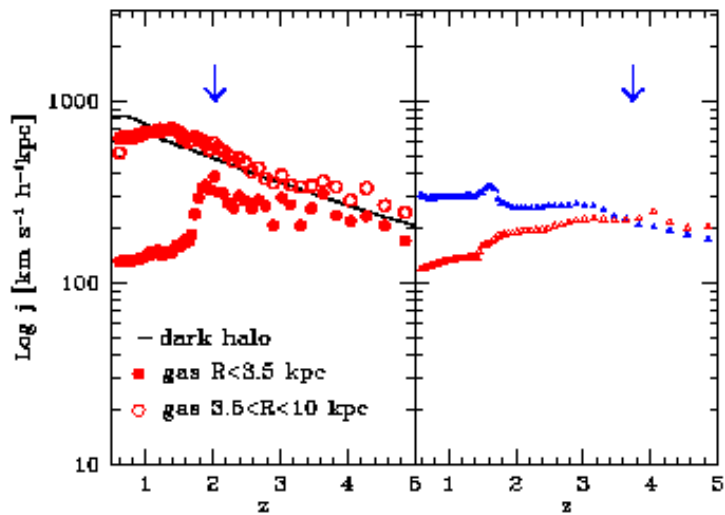}
\end{center}
\caption{How galaxies lose their angular momentum, during $\Lambda$CDM simulations: the left panel shows 
the specific angular momentum as a function of redshift of the dark matter (solid line) and the gas in the compact 
central disk (filled circles) or the gas in the envelope (empty circles). The right panel shows the evolution of 
the angular momentum of the dark matter (blue curve) or the gas (red curve) of the massive progenitor that 
has already formed a compact disk at z=2, from  D'Onghia et al. (2006).}
\label{fig3}
\end{figure}

External gas infall could help to solve the problem of
the angular momentum catastrophy, and the formation
of too small galaxy disks in the conventional scenario
(Navarro \& Steinmetz, 2000). The gas loses its angular momentum
essentially through mergers, but not during cold accretion 
(D'Onghia et al. 2006). Tidal torques and the corresponding dynamical
friction during galaxy interactions transfer the gas orbital 
angular momentum to the more extended dark matter
(e.g. Barnes \& Hernquist 1992), but also to
the extended gas in the outer filaments (see Figure \ref{fig3}).
After a violent merger, the gas has condensed in a compact disk, of
radius lower than 3.5 kpc.
Only diffuse cold accretion after the last merger can reform
an extended disk, but only a small fraction of the baryonic mass 
is then in the extended disk (larger than 3.5 kpc). 
This gas accretion can be done
either by smooth gas, or substructures less than
1/10th in mass, which are easily stripped.

Since each major merger exchanges angular momentum between the baryons and dark
matter, the latter could be expected to possess more specific angular momentum,
in haloes having experienced more mergers. In fact, this does not appear to be the case in numerical simulations, 
since after a certain period of relaxation,
the excess of angular momentum is expelled from the halo through the virial radius (D'Onghia \& Navarro 2007). 
The specific angular momentum is a function of radius, and its excess yields the expansion of the particles, 
and inflation of the haloes. 

\bigskip

{\bf Why the angular momentum problem is still not solved}

To keep a significant angular momentum in baryonic disks, there should be cold gas accretion along 
cosmic filaments after the last major merger,   where occurs the angular momentum loss.
 To form substantial disks, there should be no major mergers from z=1 until the present time,
but this is not likely, in the hierarchical scenario, and according to cosmological simulations.
Statistically, there exists dark haloes without any merger after z=1, and subsequent gas accretion
will then explain extended gas disks around big bulges. But the main problem is to form
spiral disks, without any bulge at z=0.

\bigskip

The gas inflow from filaments has an angular momentum direction generally offset
from that of the galaxy disks. This skewed accretion will in general create warps,
and progressively re-orient the spins of galaxies (e.g. Jiang \& Binney 1999).
When the offset is extreme, and the accretion is perpendicular, then polar rings
could be formed (Maccio, Moore \& Stadel 2006). This is an efficient way
to form such systems, which reveal young disks around early-type systems.
In polar ring galaxies formed in such accretions, the
dark matter component is predicted to be quite round, at least
in the region of the visible galaxy.

\section{Infall of gas and galaxy dynamics}

\subsection{Star formation history in spirals}

In order to reproduce the observed abundances of metals,
models of the chemical evolution of the Milky Way
require a continuous infall of gas with metallicity
about 0.1 times the solar value. Infall is required in particular
to solve the well-known G-dwarf problem, i.e. the observation of a very 
narrow range in metallicity for most long-lived stars near the Sun. 
The infall of gas is also supported by the constant or increasing star 
formation rate (SFR) scenario inferred from the local distribution 
of stars (e.g. Haywood, Robin \& Creze 1997). Other abundance
 problems require also an infall rate integrated over the entire disk of
the Milky Way of a few solar mass per year at least (Casuso \& Beckman 2001).
 This infall dilutes the enrichment arising from the production of heavy
 elements in stars, and thereby prevents the metallicity of the
interstellar medium from increasing steadily with time. Some of this gas
 could come from the High Velocity Clouds (HVC) infalling onto our galaxy
disk (Wakker et al. 1999).

In M31, HST colour-magnitude diagrams reveal that there exists a 
similar G-dwarf problem, even more severe than in the Milky Way
(Worthey \& Espana 2004). 
Here also closed-box models cannot explain the stellar distributions.
The star formation rate in spirals in the middle of the Hubble sequence
has also kept of the same order of magnitude over the Hubble time,
instead of the expected exponential decrease expected from closed box
models (e.g. Kennicutt et al. 1994), and favors gas accretion.

\subsection{Secular evolution}

Gas accretion is an essential ingredient for 
secular evolution driven by bars and spirals.
Without gas, stellar disks are heated by spiral
waves, and  become featureless, or only barred
after several dynamical times.
  Gas accretion can maintain spiral structure,
and also can reform bars, which are weakened or
destroyed in the evolution.  Bar/spiral gravity
torques produce radial gas inflow, and nuclear starbursts.
Dynamical instabilities then regulate themselves, since the
gas inflow itself can destroy the bar. The bar destruction involves
two mechanisms: first the central mass concentration built after the
gas inflow, destroys the orbital structure sustaining the bar,
scatter particles and push them on chaotic orbits. 
Second, the gas inflow itself weakens
the bar, by exchanging its angular momentum to the stars
forming the bar (Bournaud, Combes \& Semelin 2005). 

Secular evolution then includes several bar episodes in a galaxy
life-time. A spiral galaxy rich in gas (at least 5\% of the disk mass)
is unstable with respect to bar formation. Gravity torques are then
efficient to drive the matter inwards. The galaxy morphological type
evolves towards early-types, the mass is concentrated, the bulge
is developped, through horizontal and vertical resonances.
This weakens the bar, and when the galaxy becomes again
axi-symmetric, gas can be accreted from the outer parts by
viscosity (Bournaud \& Combes 2002). The gas accretion,
if significant with respect to the disk mass, can reduce
the bulge-to-disk ratio (by replenishing the disk), and
make the galaxy disk unstable again to a new bar.

\subsection{AGN fueling}

Gas flows due to bar gravity torques are an attractive explanation
to explain AGN fueling. However, low-luminosity AGN, such as Seyfert galaxies
are not observed in good correlation with bar strength. This
could be due to the widely different time-scales between the 
gas flows from the outer parts of galaxies to the nuclear region (100pc-1kpc)
and the nuclear short time-scales. Also, the fact that gas flows destroy
the bar could play a large role in this absence of correlation.

More details in the evolutionary scenario have been searched
through the gas mapping at high resolution in a dozen of Seyfert
galaxies by the NUGA project (e.g. Garcia-Burillo et al. 2005).
In most galaxies, gas has been found concentrated in rings at
inner Lindblad resonances (ILRs), and the nuclear torques were
measured positive, meaning that no fueling was occuring.

The accumulation of the gas into ILR rings is the first step of
the secular evolution scenario, in galaxies where mass is concentrated in bulges, 
which is likely for spiral galaxies with massive central black holes. Through bar torques,
the gas is driven inwards from corotation to the ILR, and accumulates in the ring, usually the site 
of active star formation. Since the sign of the torques change at resonance, 
positive gravity torques then prevent the gas to flow further
in: under the bar forcing, gas inside the ring is evacuated outwards. 
Only in the central 10pc around the black hole, the potential is
mostly axi-symmetric, and viscosity could help to fuel the AGN.
A more consequent feeding has to wait the weakening of the bar,
through the gas flows. Then the gas of the
nuclear ring is liberated to form a smoother disk through viscosity which has become
competitive against gravity torques. 

The AGN fueling in early-type spirals can be viewed as a two step process:
first gravity torques bring the gas of the large-scale disk to the nuclear ring, and when the
dynamical feedback has weakened the bar, the viscous torques smooth out the ring, and bring
gas to the central 10~pc, where it is under the influence of the keplerian potential of the black
hole. The disk becomes axi-symmetric and the cycle can be restarted at its first step. In
particular, the disk will be prone to a new bar instability if gas is accreted from the outer parts
of the disk.  In any case, the AGN phase is always correlated with a weak bar phase.

\begin{figure}[ht]
\begin{center}
\includegraphics[angle=-90,width=10cm]{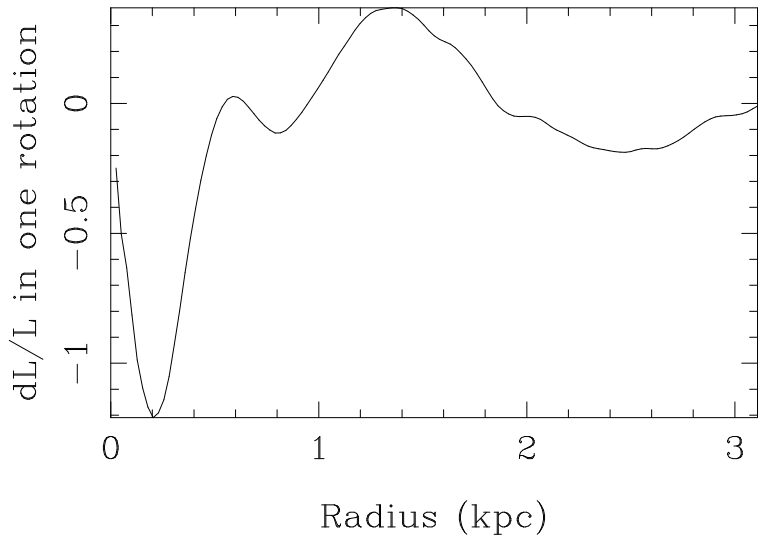}
\end{center}
\caption{Estimation of the gravity torques, and the relative angular momentum
 lost by the gas in a rotation period (L is the angular momentum
perpendicular to the disk). The torque is negative towards the center, which means the gas is 
fueling the AGN in NGC 2782 (Hunt et al. 2007).}
\label{fig4}
\end{figure}

Only in one galaxy, NGC 2782 (Hunt et al. 2007), the nuclear torques have
been measured negative, and the fueling caught in action (see Fig \ref{fig4}). 
 This means that the fueling phase is very short, and only a minor
fraction of active galaxies are observed in this phase.

\section{Cooling flows in galaxy clusters}

Galaxy clusters are interesting to study in the 
context of gas accretion, since they are certainly the
prototype model of " hot accretion " in massive dark matter haloes.
Since the hot gas is observed, and contains most of
the baryons, it can be sure that cold accretion is
restricted to a minimum.
The cooling time of the gas at the center of clusters
becomes smaller than the Hubble time, so an important
gas flow has been predicted since many years. Now the
cold gas has been observed (23 detected galaxies in CO at 2.6mm,
 Edge 2001; Salom\'e \& Combes 2003), but 
the amplitude of the flow is 10 times lower than expected in the early times.
Moreover, X-ray lines reveal gas down to 
a temperature of 2kev but not below (Peterson \& Fabian 2006).

The last decade results from the Chandra \& XMM satellites
have established that the cooling flow phenomenon is self-regulated.
The simple view of spherical unimpeded flow is now
completely revised, to take into account the
re-heating processes, the feedback due to the active nucleus or black
hole: shocks, radio jets, acoustic waves, bubbles...
The cooling gas is not only observed in the CO emission,
but also directly from the pure rotational lines of the H$_2$
molecule. For instance, large molecular masses are infalling
onto the central galaxy in Perseus (Johnstone et al. 2007).
 The H$_2$ lines have been surprisingly strong in many
cooling flow clusters, as discovered by Spitzer (e.g.
Egami et al. 2006).

\subsection{Amount of cooling gas}

One of the puzzle in cooling flows was
that the gas between temperature 10$^4$ and 10$^7$K was not seen with the 
expected amount. Gas at T = 10$^{5.5}$K is now seen in the  
OVI emission (with FUSE) in many clusters,
for instance in A1795 and Perseus (Bregman et al. 2006), giving cooling 
rates of $\sim$30 M$_\odot$/yr, about 3 times lower
than previously expected.

The cold gas masses detected through CO emission are compatible
with these cooling rates. The masses derived are conservative,
since obtained with the same CO-to-H$_2$ conversion factor
as solar metallicity spiral galaxies. Masses are
between 3 10$^8$ and 4 10$^{10}$ M$_\odot$ 
(Edge 2001; Salom\'e \& Combes 2003). 
In some cases, high resolution maps have been
performed with millimeter interferometers, and it
was clearly established that the CO emission is
not correlated with any central galaxy, but is originating
from the cooling flow. The CO gas kinematics is related to the cluster
and is not rotation around the central galaxy. This is the
case for Abell 1795, where a cooling wake
is observed in X-ray (Fabian et al. 2001). 
The cooling rate has been estimated to  
200 M$_\odot$/yr inside a radius of 200kpc (Ettori et al. 2002).
The CO emission has been shown to follow the cooling wake
(Salom\'e \& Combes 2004), and also the 60kpc H$\alpha$ filament mapped by
Cowie et al. (1983). The main CO velocity is centered
on the cluster velocity. In Perseus, the CO emission follows
the conspicuous H$\alpha$ filamentery structure, mapped by 
Conselice, Gallagher \& Wyse (2001). Again the molecular 
component is not in rotation in NGC 1275 (Salom\'e et al. 2006).
The radio jets from 3C84 create holes and bubbles in the hot gas, 
which is compressed at the boundaries, and cools faster.
This means that cold clumps could be found far from the central
cooling activity.

Two main models are therefore advanced to account
for cooling and heating processes: hot and cold feedback.
In the first point of view of hot feedback, the hot ICM  
cools and is accreted close to the AGN. Then the 
observed cold gas far from the center should come from galaxies.
In the cold feedback scenario, cold gas is accreted on large scales.
Cooling can occur at large distance
because of non-linear over-dense blobs, created in wide
regions due to past AGN activity, which decrease the cooling time-scale
for these perturbed regions (e.g. Pizzolato \& Soker 2005).
In this point of view, there is co-existence of multi-phase
(cold and hot) gas.

\subsection{Numerical simulations}

The cold feedback scenario has been considered in recent
multi-phase simulations of the cluster environment
 (Revaz, Combes \& Salom\'e 2007).
The AGN feedback is schematized by following hot bubbles,
during their buoyant rise across the cluster.
Very cold gas is spontaneously formed by the perturbation of the hot ICM,
and forms a filamentary structure in the wake and in the rim of the bubbles.
The amount of cold gas formed and its kinematics are in agreement
with observations of CO and H$\alpha$.

The AGN feedback is therefore not only negative in re-heating
the cooling gas, but also positive in allowing some further gas
to cool. 

\begin{figure}[ht]
\begin{center}
\includegraphics[width=6cm]{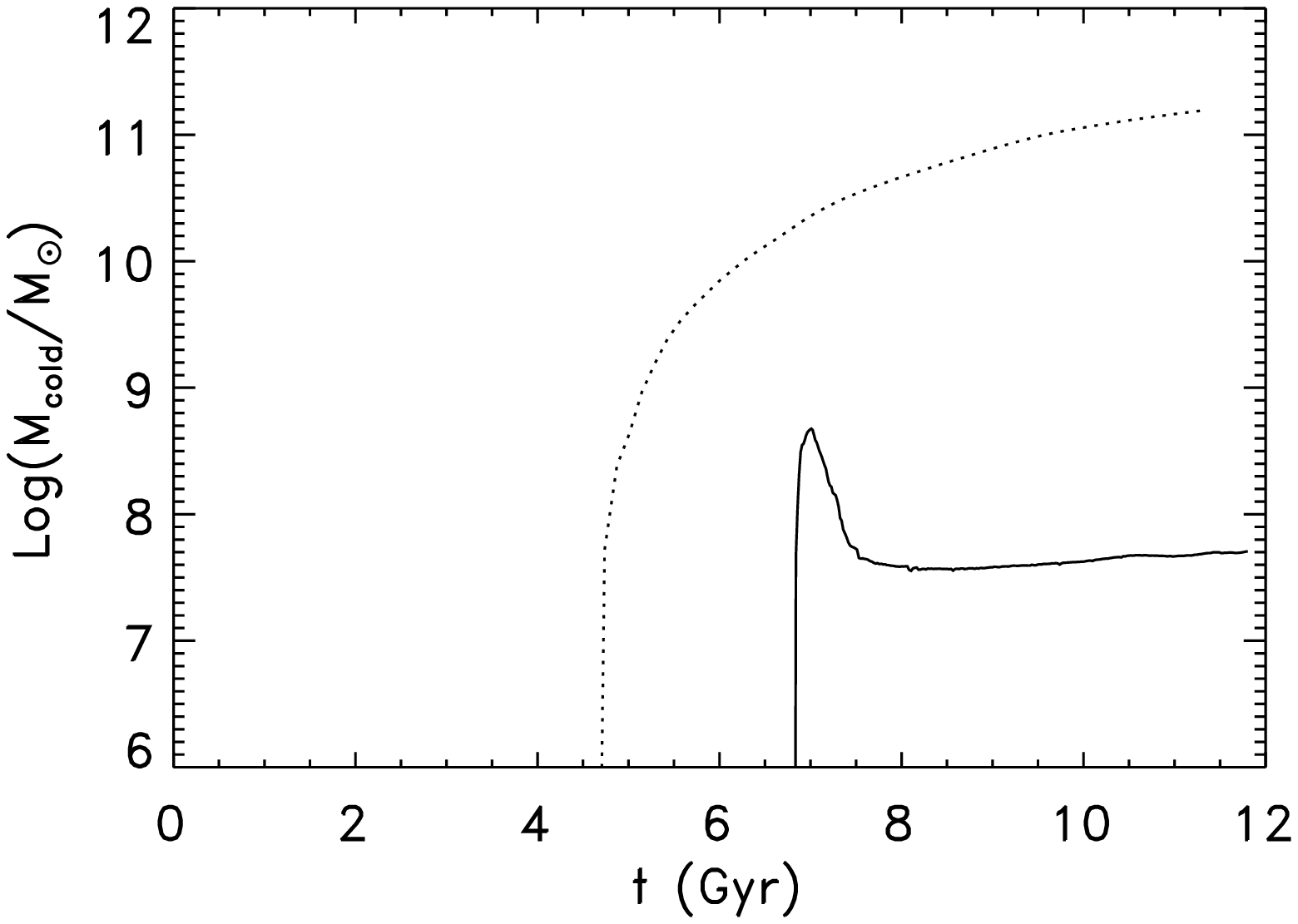}
\includegraphics[width=6cm]{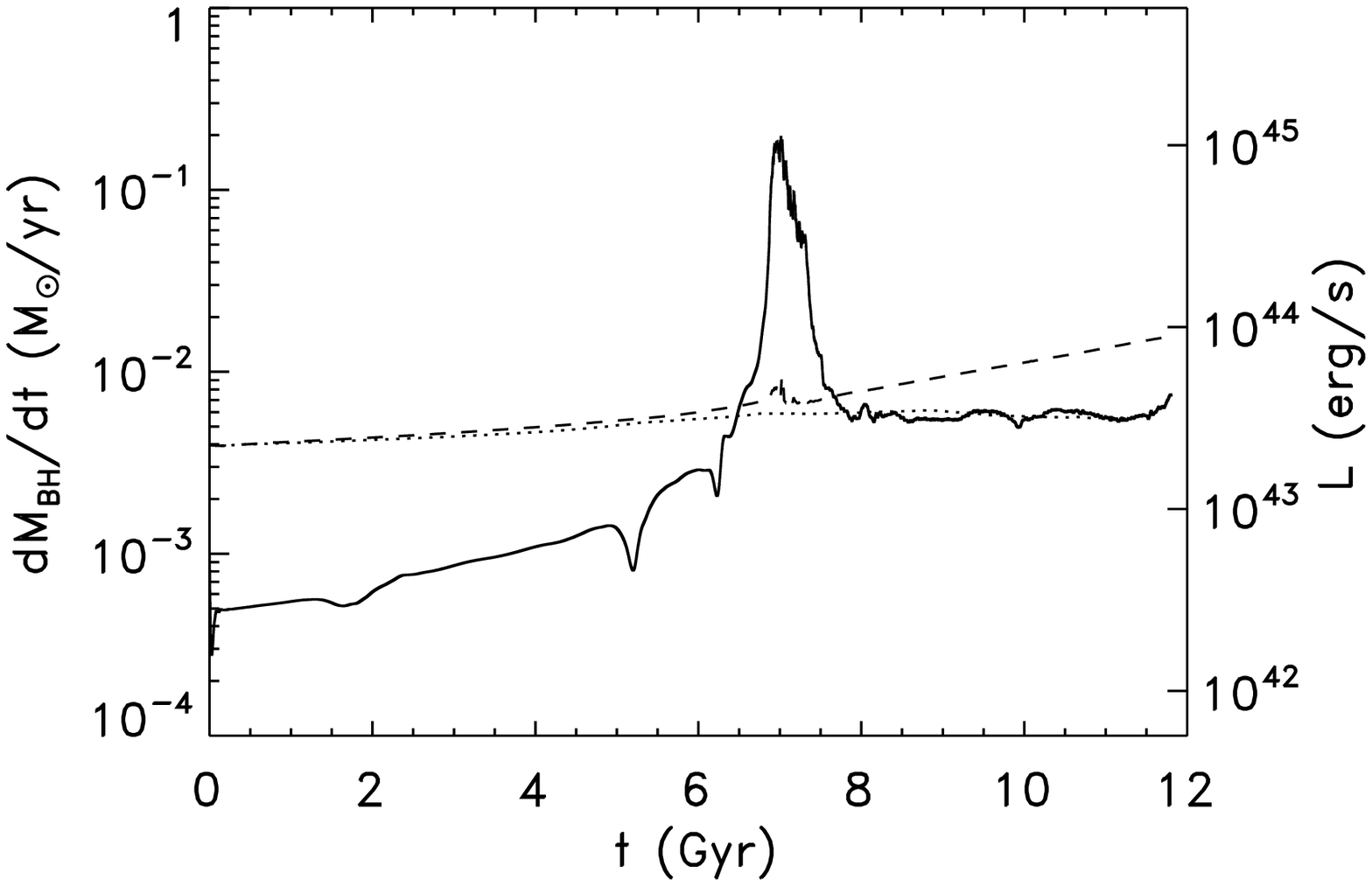}
\end{center}
\caption{AGN feedback in cooling flows. The left panel shows the mass
of the cold gas as a function of time, in the cooling flow simulation
(dotted line), and with AGN feedback (solid line); the solid line in the
right panel shows the accretion rate by the black hole (left axis) or the 
corresponding jet mechanical luminosity (right axis) in the simulation
with AGN feedback. The dashed line shows the X-ray emitted by the ICM
in the same simulation, compared with the dotted line without
feedback, from Cattaneo \& Teyssier (2007).}

\label{fig5}
\end{figure}

Alternatively, the hot feedback scenario has been also 
simulated by Cattaneo \& Teyssier (2007).
They show that the catastrophic cooling is 
regulated by the AGN jets, and a stationary state
can be reached, to reproduce observations (see Fig \ref{fig5}).
The cooling gas is regulated to just the right amount 
required to fuel the AGN activity. 

\section{Conclusions}

Cold gas accretion along cosmic filaments is an important
way to form and grow galactic disks. In particular cold gas accretion
dominates at high redshift. 
Simulations with enough resolution reveal that all the gas is not
shock heated at the virial radius, as in the usually adopted paradigm.

The cold accretion is a way to solve the angular momentum
problem in the formation of galaxy disks. In the hierarchical
scenario, the angular momentum of baryons is lost in mergers
to the dark matter haloes. The gas accreted after the last merger
does not lose angular momentum, and is able to form large
disks, although around already massive bulges.
 The baryons will keep more specific angular momentum globally,
if galaxy formation does rely more on smooth matter accretion than mergers.

Gas accretion is also required for disk evolution, being
one of the main actor of secular evolution. It is needed
to maintain spiral structure, reform bars, 
asymmetries, warps..

Cooling flows in galaxy clusters are a prototype
example of hot accretion.  Cold gas observed in CO
and H$_2$ line emission is now compatible with the 
expected revised cooling rates. The cooling process
is self-regulated through AGN feedback. The hot feedback 
scenario considers that the self-regulation can occur
close to the central AGN, while cold feedback involves
a much wider region, where dense clumps of gas can cool
far from the AGN.  This scenario could explain the 
large-scale filamentary structure observed in CO and
H$\alpha$ in nearby cooling flows.

\acknowledgements 
 I am very grateful to the organisers for a very instructive and pleasant meeting, that reflected 
so vividly John Beckman's eclectic career.

\end{document}